\documentclass[prb,twocolumn,longbibliography]{revtex4-1}
\usepackage{graphicx}
\usepackage{bm}
\usepackage{wrapfig}
\usepackage{color}
\usepackage{natbib}
\usepackage{hyperref}
\usepackage{amssymb}
\usepackage{amsmath}
\usepackage{geometry}
\usepackage{comment}

\def\XXint#1#2#3{{\setbox0=\hbox{$#1{#2#3}{\int}$}
     \vcenter{\hbox{$#2#3$}}\kern-.5\wd0}}

\begin{document}

\title{Electronic scattering off a magnetic hopfion}

\author{Sergey S. Pershoguba, Domenico Andreoli, and Jiadong Zang}

\affiliation{Department of Physics and Astronomy, University of New Hampshire, Durham, New Hampshire 03824, USA}

\date{\today}

\begin{abstract}
We study scattering of itinerant electrons off a magnetic hopfion in a three-dimensional metallic magnet described by a magnetization vector $\bm S(\bm r)$. A hopfion is a confined topological soliton of $\bm S(\bm r)$ characterized by an {\it emergent} magnetic field $B_\gamma(\bm r) \equiv \epsilon_{\alpha\beta\gamma} \,\bm S\cdot(\nabla_\alpha \bm S\times \nabla_\beta \bm S)/4 \neq 0$ with vanishing average value $\langle \bm B(\bm r)\rangle = 0$. We evaluate the scattering amplitude in the opposite limits of large and small hopfion radius $R$ using the eikonal and Born approximations, respectively. In both limits, we find that the scattering cross-section contains a skew-scattering component giving rise to the Hall effect within a hopfion plane. That conclusion contests the popular notion that the topological Hall effect in non-collinear magnetic structures necessarily implies $\langle \bm B(\bm r)\rangle \neq 0$. In the limit of small hopfion radius $pR \ll 1$, we expand the Born series in powers of momentum $p$ and identify different expansion terms corresponding to the hopfion anisotropy, toroidal moment, and skew-scattering. 
\end{abstract}

\maketitle

\section{Introduction}

In the celebrated paper \cite{AB_paper}, Aharonov and Bohm considered scattering of electrons off a solenoid carrying magnetic flux $\Phi$ and showed that the differential cross-section is a periodic function of $\Phi$. That work  laid the foundation for the discussion of the topological effects in quantum mechanics. In many respects, the recent investigation on the topological Hall effect~\cite{Back_2020,gobel21} in non-collinear magnetic textures is the most recent incarnation of the Aharonov-Bohm physics. In the appropriate transport regime \cite{Denisov2020}, the non-collinear spin configuration generates a (fictitious magnetic) field\cite{Tesanovic1999} $B_\gamma(\bm r) \equiv \epsilon_{\alpha\beta\gamma} \,\bm S\cdot(\nabla_\alpha \bm S\times \nabla_\beta \bm S)/4 \neq 0$, which produces a skew-scattering deflection of carriers.  For example, a magnetic skyrmion, observed in two-dimensional magnetic films \cite{skyrmion_observation2009,skyrmion_observation2010,skyrmion_observation2011}, generates a fictitious magnetic flux equivalent to the flux quantum. Therefore, electronic scattering  off such structures closely resembles the Aharonov-Bohm set up. Owing to a small size (large density) of skyrmions,  the fictitious magnetic field $B$ produced in such structures may be an order of magnitude larger ($\sim 500$~T) than that attainable in conventional magnetic experiments ($\sim 50$~T). That magnetic field may produce a large topological Hall effect \cite{Tokura2019}. We note that the topological Hall effect was also predicted in systems without skyrmions \cite{Wen-Tao2017,Nagaosa2018chirality}.   

In recent past, there has also been a significant push to extend the research of non-collinear  magnetic structures to three dimensions (3D). Magnetic simulations \cite{Back_2020,gobel21} reveal that, under appropriate conditions, three-dimensional magnets may host a zoo of exotic magnetic textures and quasiparticles interesting from both  fundamental and practical standpoints. New experimental imaging tools~\cite{Donnelly2017} are becoming available, which may facilitate the search and identification of such objects. In this paper, we focus on one such paradigmatic topological object  - a   magnetic hopfion. Conceived originally in the context of field theory \cite{wilczek1983,Fadeev1997}, hopfions are now discussed in the realm of magnetic systems~\cite{Jiadong2018,sutcliffe,Smalyukh2018,Rybakov2019,Zang2020}. Various recipes have been proposed how to stabilize hopfions in specific materials \cite{Rybakov2019} and finite geometries \cite{Jiadong2018,sutcliffe,Smalyukh2018}. Reference~[\onlinecite{hopfion_experiment}] reported a first observation of a hopfion in a magnetic nano-disk. Hopfions were also discussed in the context of superconducting \cite{Babaev2002} and ferroelectric systems\cite{hopfion_ferroelectric}.

\begin{figure}
	(a)\includegraphics[width=0.8\linewidth]{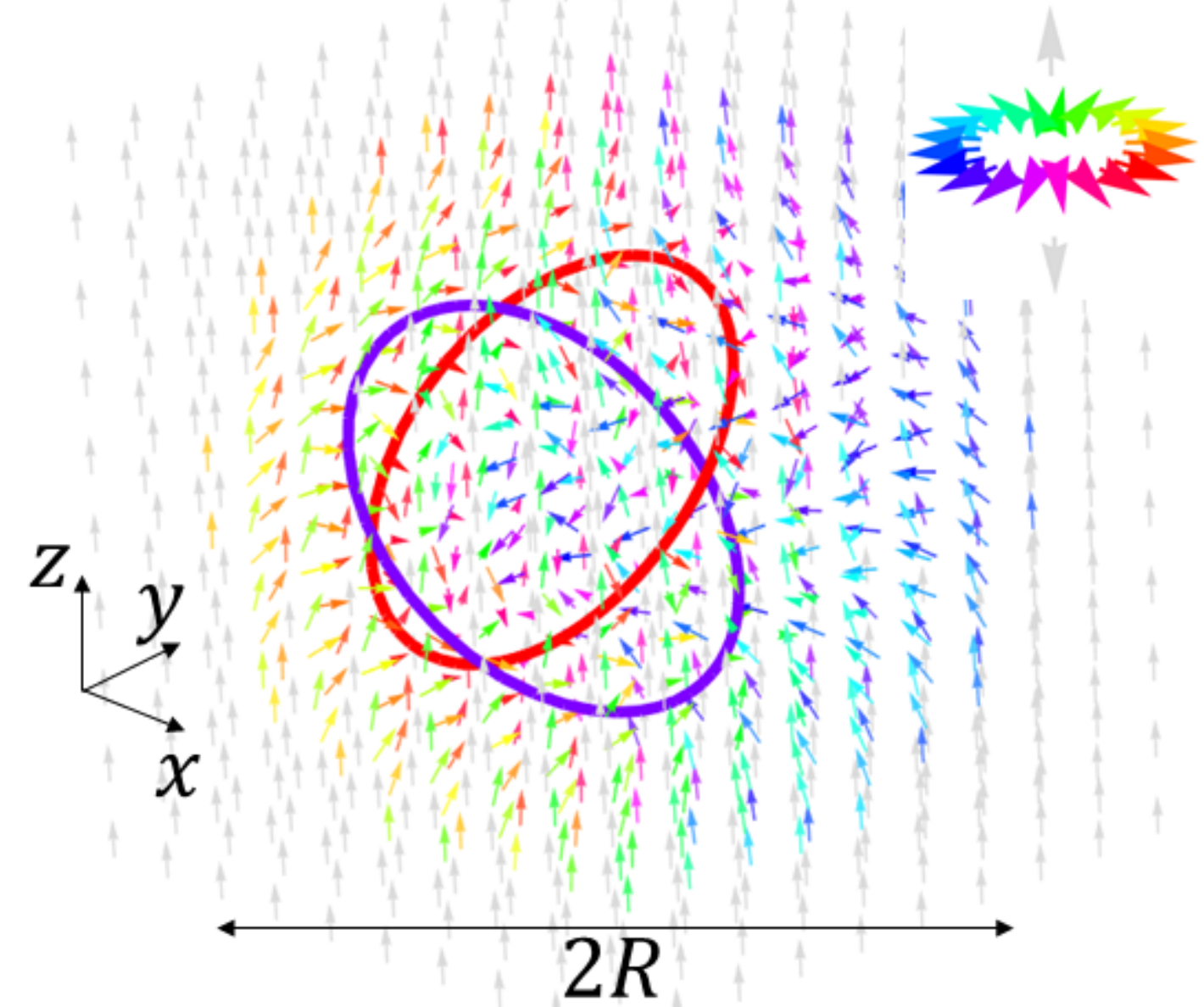} \\
	\vspace{0.5cm}
	(b)\includegraphics[width=0.95\linewidth]{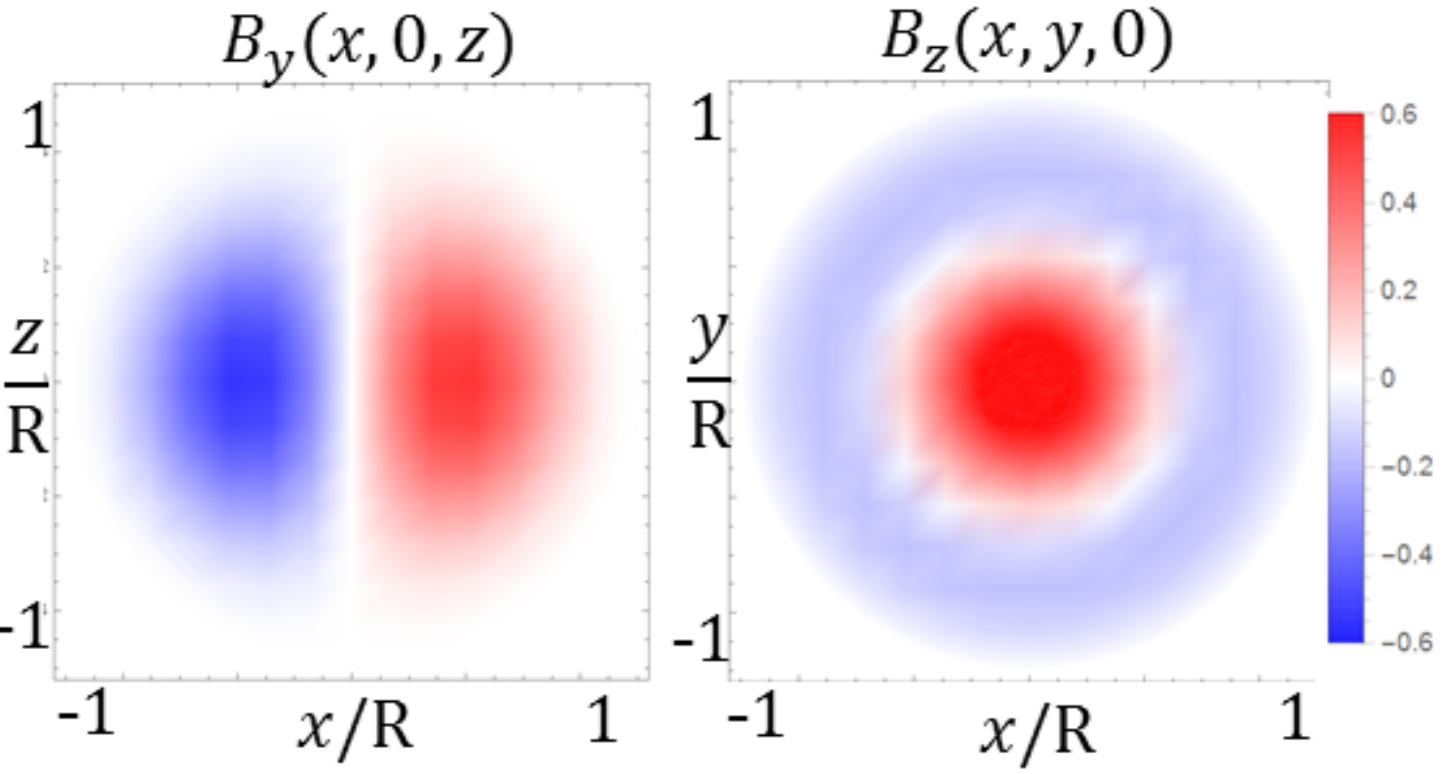}
	\caption{Hopfion texture. (a) Magnetic vector $\bm S(\bm r)$ in 3D. Coloring scheme is shown in the top-right color: the vectors with components in the $z = 0$ plane are shown in colors, whereas vectors with $\bm S(\bm r) \parallel \hat{\bm z}$ are shown in gray. The two linked contours are the solutions of equations $\bm S(\bm r) = \hat{\bm x}$ (red) and $\bm S(\bm r) = -\hat{\bm y}$ (blue). (b) Profile of the emergent magnetic field $\bm B(\bm r)$ in the  $y=0$ plane (left) and  $z=0$ plane (right). Average field vanishes, see Eq.~(\ref{beq0}).} 
	\label{fig:hopfion}
\end{figure}

It is an appropriate point to mention that a hopfion has a non-trivial profile of the emergent magnetic field $\bm B(\bm r)$ [see Fig.~\ref{fig:hopfion}(b)]. It is characterized by a non-vanishing Hopf number 
\begin{align}
    Q = \frac{1}{(2\pi)^2}\int d^3r\, \bm B(\bm r)\cdot \bm A(\bm r), \label{hopf_index}
\end{align}
where $\bm A(\bm r)$ is the associated vector potential, i.e. $\bm B(\bm r) = \bm \nabla \times \bm A(\bm r)$. Another notable feature is that the average emergent magnetic field vanishes
\begin{align}
    \langle \bm B(\bm r) \rangle \equiv \int d^3r \,
    \bm B(\bm r) = 0. \label{beq0}
\end{align}
Nevertheless, as we show in this work, a hopfion configuration does lead to skew-scattering and the Hall effect within the {\it hopfion plane}.

In this paper, motivated by the original Aharonov-Bohm paper\cite{AB_paper} as well as the recent interest in 3D magnetic systems, we pose a hitherto unexplored question of electronic scattering off a topological magnetic object in 3D. We consider a hopfion configuration in a metallic magnet and evaluate the scattering amplitude of the itinerant electrons. We lay out the basics of the hopfion geometry in Sec.~\ref{sec:hopfion_basics} and proceed to evaluating the scattering amplitude in Sec.~\ref{sec:scattering}. A challenge is that a hopfion does not have spherical symmetry, so a standard method of expanding in partial wave harmonics is not applicable. Therefore, in order to evaluate to the scattering amplitude, we resort to the eikonal and Born approximations for large $pR \gg 1$ and small $pR \ll 1$, respectively. Here, $R$ and $p$ are the hopfion radius and electronic momentum, respectively. We provide a detailed account of the applicability of these approximations in Sec.~\ref{sec:applicability}. In Sec.~\ref{sec:eikonal}, we proceed to evaluating the scattering amplitude in the eikonal approximation. We find that the differential cross-section contains a skew-asymmetric component within a {\it hopfion plane}, but the average transferred momentum in the transverse direction vanishes, i.e. $\langle \Delta \bm p_{transverse} \rangle$. However, as we explain, the latter exact equality is an artifact of the eikonal approximation. There are two implicit assumptions ``under the hood'' of the eikonal approximation: (i) that the semiclassical approximation is applicable and (ii) that semiclassical trajectories may be approximated as straight lines. Departing from either of the two conditions renders $\langle \Delta \bm p_{transverse} \rangle \neq 0$, and, hence, the associated Hall effect to survive as well. That conclusion contests the widespread belief in the community\cite{Kanazawa2015,Ohuchi2015,Yun2018} that non-zero topological Hall signal necessarily implies $\langle \bm B \rangle \equiv \int dr \bm B(\bm r) \neq 0$. In the deeply quantum regime $pR\ll 1$, we evaluate the scattering amplitude using the Born series (truncated to second order) in Sec.~\ref{sec:born}. As a bi-product, we also evaluate the second-order Born approximation for a Gaussian-type potential in Appendices~\ref{sec:born_appendix} and \ref{sec:integral_appendix}. To the best of our knowledge, only a similar calculation for the Yukawa potential exists so far \cite{GalitskiBook}. That calculation allows to expand the scattering amplitude in powers of momentum. Different terms in that expansion correspond to the hopfion anisotropy, toroidal moment of the hopfion, skew-scattering contributions, etc.  We comment on the possible manifestation of those terms in transport in Sec.~\ref{sec:transport} and offer concluding remarks in Sec.~\ref{sec:conclusion}.

\section{Magnetic hopfion.} \label{sec:hopfion_basics}
To set the stage, we discuss details of a hopfion texture in this section. We consider a 3D ferromagnet described by a magnetization vector $\bm S(\bm r)$ normalized to unity $|\bm S(\bm r)| = 1$. A hopfion is a localized topological soliton of the field $\bm S(\bm r)$. We use the following parametrization of the hopfion~\cite{Zang2020} 
\begin{align}
     \bm S(\bm r) &= \hat{\bm z} + \delta \bm S(\bm r), \label{hopfion}\\
    \delta\bm S(\bm r)  &=\frac{\sin 2\eta(r)}{r}\left(\begin{array}{c}
         x  \\
         y \\
         0
\end{array}\right) - \frac{2\sin^2\eta(r)}{r^2}\left(\begin{array}{c}
             -yz \\
              xz \\
              x^2+y^2
\end{array}\right).  \nonumber
\end{align}
Here, $\hat{\bm z}$ describes a uniform magnetization at $r\to \infty$, whereas $\delta \bm S(\bm r)$ encapsulates the localized hopfion texture. The phase $\eta(r)$ is an arbitrary monotonic function of $r = \sqrt{x^2+y^2+z^2}$ with constraints  $\eta(0) = 0$ and $\eta(\infty) = \pi$. It controls the extent of the hopfion in the radial direction.
The texture~(\ref{hopfion}) has cylindrical symmetry around $\hat{\bm z}$ axis. For that reason, $\hat{\bm z}$ axis is referred to as the {\it hopfion axis}, and  $z = 0$ - the {\it hopfion plane}. A hopfion occupies finite space, as illustrated in Fig.~\ref{fig:hopfion}(a), and may be thought of as a localized magnetic quasiparticle. Its dynamics under the applied electric current was studied in Ref.~[\onlinecite{Zang2020}].

A complementary description of a hopfion may be obtained by evaluating an emergent field ${B_\gamma(\bm r) = \epsilon_{\alpha\beta\gamma} \,\bm S\cdot(\nabla_\alpha \bm S\times \nabla_\beta \bm S)/4}$. We evaluate both the field
\begin{align}
    \bm B = & \frac{2\cos\theta \sin^2\eta(r)}{r^2}\,\bm e_{r}- \frac{\sin\theta\,\sin 2\eta(r)\,\eta'(r)}{r}\,\bm e_{\theta} \nonumber \\
    & \qquad\qquad\qquad + \frac{2 \sin\theta \sin^2\eta(r)\,\eta'(r)}{r}\,\bm e_{\phi} \label{B} 
\end{align}
and the associated vector potential 
\begin{align}
    \bm A = 2 \cos\theta \sin^2\eta(r)\,\eta'(r)\,\bm e_r + \frac{\sin\theta\,\sin^2\eta(r) }{r}\,\bm e_\phi \label{A}
\end{align}
satisfying the conventional relation $\bm B = \bm\nabla\times\bm  A$. In writing Eqs.~(\ref{B}) and (\ref{A}), we used spherical coordinates, where $\cos\theta = z/r$ and $\bm e_r$, $\bm e_{\theta}$, $\bm e_{\phi}$ denote the orthogonal unit vectors in the radial, polar and azimuthal directions with respect to the {\it hopfion axis} $\hat{\bm z}$. 

The topological character of a hopfion may be illustrated in two complementary ways: either directly from the spin-configuration $\bm S(\bm r)$ or using the Hopf number $Q$. To illustrate the former, let us pick arbitrary two vectors on the unit sphere $\bm S_1$ and $\bm S_2$, i.e. $|\bm S_1|= |\bm S_2| = 1$. Then, the two contours, determined by the solutions of the equations $\bm S(\bm r) = \bm S_1$ and $\bm S(\bm r) = \bm S_2$, are linked. A specific example, corresponding to $\bm S_1  = \hat {\bm x}$ and $\bm S_2 = \hat {\bm y}$, is shown in Fig.~\ref{fig:hopfion}(a). On the other hand, the linking number between these contours equals \cite{Nicole_1978} the topological Hopf number $Q$ defined in Eq.~(\ref{hopf_index}). We substitute Eqs.~(\ref{B}) and (\ref{A}) in Eq.~(\ref{hopf_index}) and verify the value $Q = 1$. 

We note that the average emergent magnetic field vanishes according to  Eq.~(\ref{beq0}). The profile of the field in the $y=0$ and $z=0$ planes is shown in the left and right panels of Fig.~\ref{fig:hopfion}(b). The field in the $y=0$ plane has a skyrmion-antiskyrmion structure. The field in the hopfion plane $z = 0$ has a structure reminiscent of a target skyrmion\cite{targetSk2017}.  As usual in electrodynamics, a non-uniform distribution of the field $\bm B(\bm r)$ may be characterized by moments. Evaluating the first-order moment $B_{\alpha,\beta} \equiv \int d^3r \,B_\alpha(\bm r) r_\beta$ for the field~(\ref{B}), we find
\begin{align}
    & B_{\alpha,\beta} =  \epsilon_{\alpha\beta\gamma}L_\gamma, \nonumber\\
    & \bm L =  \frac{1}{2}\int d^3r \, [\bm r \times \bm B(\bm r)] = L \,\hat {\bm z}, \label{toroidal}
\end{align}
where $L = \frac{2}{3}\int d^3r\,\sin^2[\eta(r)]\,\eta'(r)$. Vector $\bm L$ is referred to as the toroidal moment \cite{Dubovik1990} and originates from the azimuthal component ($\propto \bm e_{\phi}$) of the field $\bm B(\bm r)$ winding along a torus. A slice of that torus is shown in the left panel of Fig.~\ref{fig:hopfion}(b). 

The two representations of a hopfion either via a magnetic texture~(\ref{hopfion}) or the emergent fields~(\ref{B})-(\ref{A}) are complementary.

\section{Scattering off a hopfion} \label{sec:scattering}
In this section, we evaluate the scattering amplitude. A standard method of decomposition in spherical harmonics is not applicable because hopfion texture~(\ref{hopfion}) does not have spherical symmetry. Therefore, we resort to approximate methods: the eikonal approximation and the Born approximation. Below, in Sec.~\ref{sec:applicability}, we address the applicability of the two approximations.

\subsection{Hamiltonian and applicability conditions.} \label{sec:applicability}
\begin{figure}
	(a)\includegraphics[width=0.95\linewidth]{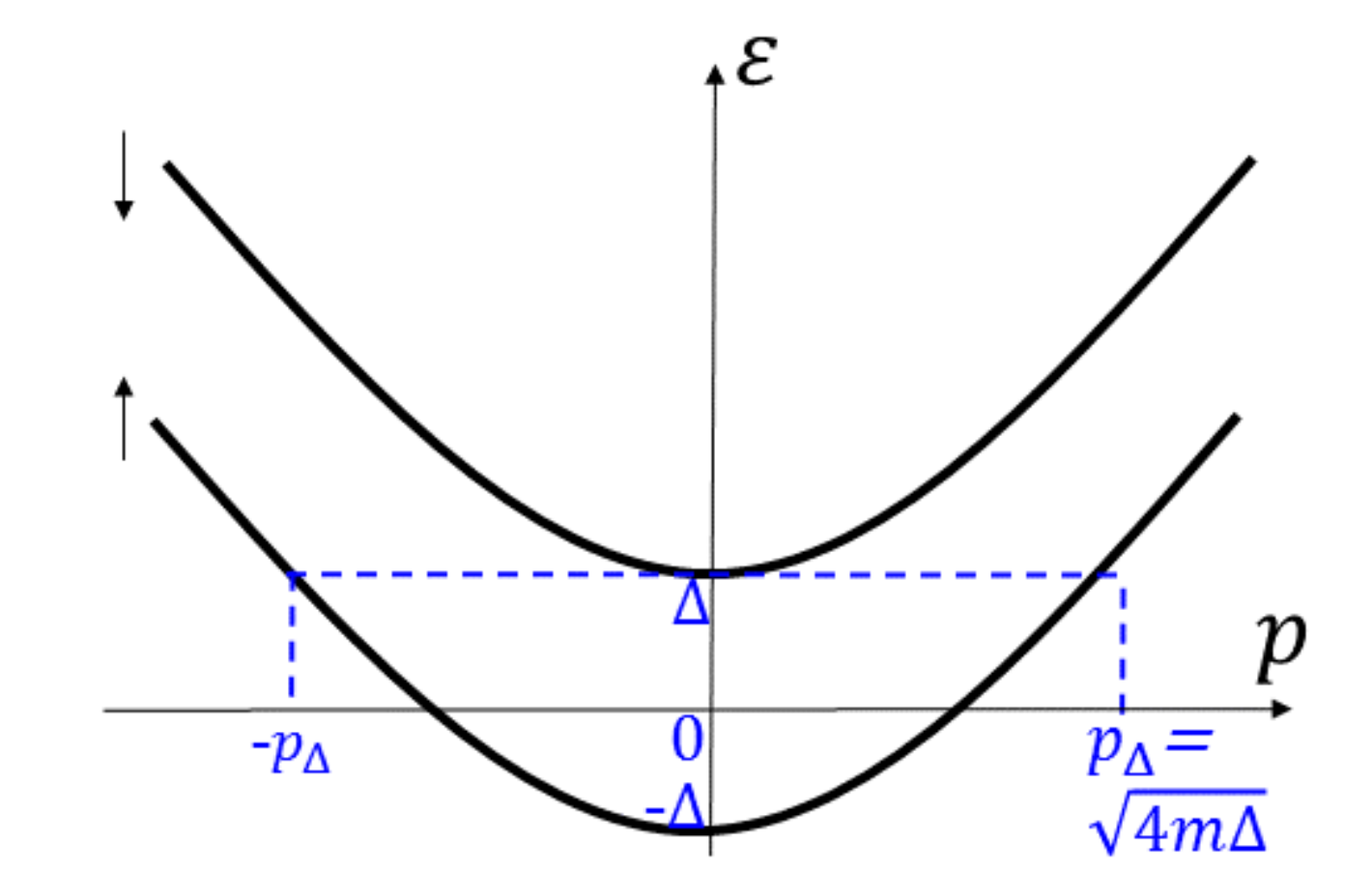} \\
	\vspace{0.5cm}
	(b)\includegraphics[width=0.95\linewidth]{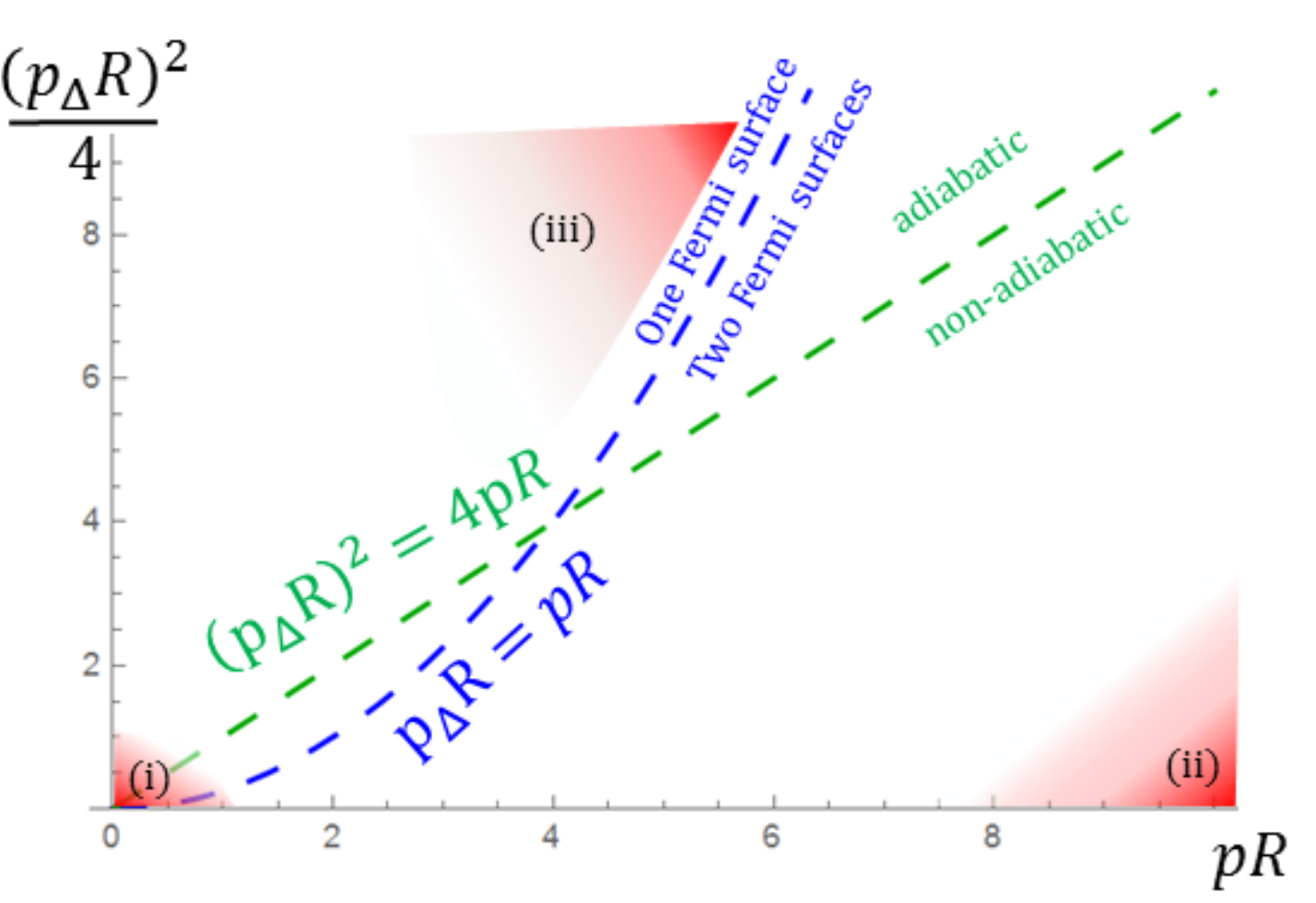}
	\caption{(a) Energy spectrum of a two-band model given by Hamiltonian~(\ref{hamiltonian}). For simplicity, we assume that only the lower band is occupied, i.e. $p<p_\Delta$. (b) Diagram of applicability of the Born and eikonal approximations. The Born approximation, analyzed in Sec.~\ref{sec:born}, is applicable in the domains (i) and (ii). The eikonal approximation is evaluated in the case of one Fermi surface in the adiabatic limit in Sec.~\ref{sec:eikonal}. The shaded area (iii) indicates the domain where it is applicable.} 
	\label{fig:applicability}
\end{figure}
Here, we define the Hamiltonian and discuss the applicability of approximations used in the following sections. We assume that the magnetic system, described by vector $\bm S(\bm r)$, is embedded in a metallic host, so the total Hamiltonian is
\begin{align}
    &H = H_0 + V, \label{hamiltonian}\\
    &H_0 = \frac{\bm p^2}{2m} -  \sigma_z \Delta-\mu, \quad V =  - \Delta\, \delta\bm S(\bm r)\cdot \bm \sigma, \nonumber
\end{align}
where $\bm p$ is a 3D momentum of itinerant electrons. The terms proportional to $\Delta$ describe the exchange coupling between the spin $\bm \sigma$ of itinerant electrons and the static magnetization vector $\bm S(\bm r)$. The Hamiltonian~(\ref{hamiltonian}) is split into the bare~$H_0$ and the perturbation~$V$ part induced by the hopfion. We use units $\hbar = 1$ throughout this work.

Following Ref.~[\onlinecite{Denisov2020}], we examine basic parameters that determine different scattering regimes in this subsection. The electronic energy spectrum of the bare Hamiltonian~$H_0$ consists of two branches $\varepsilon_{1,2} = \frac{p^2}{2m} \pm \Delta$ shifted by the energy gap $2\Delta$, as shown in Fig.~\ref{fig:applicability}(a). The electrons with energy in the interval $-\Delta<\varepsilon<\Delta$ occupy only the bottom band, whereas the electrons with higher energy $\varepsilon > \Delta$ may occupy both bands. In momentum variables $p$, the boundary between the two domains is determined by the equation $p = p_\Delta$, where $p_\Delta = \sqrt{4m\Delta}$ is the momentum associated with energy $2\Delta$. That boundary is illustrated with a dashed blue line in Fig.~\ref{fig:applicability}(b). For simplicity, we restrict the discussion thoughout this work to the case with a single Fermi surface, i.e. $p < p_\Delta$.

The dynamics of the  electronic spin is determined by the adiabaticity parameter $\lambda = \tau\Delta$. Here,  $\tau = Rm/p$ is the time it takes to traverse the hopfion, and $\Delta$ is spin precession frequency. If $\lambda \gg 1$, the electronic spin adjusts to the local magnetic direction $\bm S(\bm r)$ as an electron travels through the magnetic texture. In the opposite regime $\lambda \ll 1$, the spin does not keep up with a fast motion of the electron. The former regime is referred to as an adiabatic and the latter as non-adiabatic. It is convenient to re-write these conditions in dimensionless variables $pR$ and $(p_\Delta R)^2$ as follows $pR \ll (p_\Delta R)^2/4$ and $pR \gg (p_\Delta R)^2/4$ for the adiabatic and non-adiabatic regimes, respectively. The line separating these domains $pR = (p_\Delta R)^2/4$ is shown in green in Fig.~\ref{fig:applicability}(b).

In our work, we evaluate the scattering amplitude using the Born and eikonal approximations. Let us comment on their applicability conditions. The applicability of Born approximation\cite{LandauBook} in the long-wavelength $pR \ll 1$ and short-wavelength $pR \gg 1$ limits are $ mR^2 \Delta \equiv (p_\Delta^2 R)/4 \ll 1$ and $mR^2 \Delta \equiv (p_\Delta^2 R)/4 \ll pR$, respectively. Both  domains are shown schematically as shaded regions (i) and (ii) in Fig.~\ref{fig:applicability}(b). In Sec.~\ref{sec:born}, we evaluate the scattering amplitude in the long-wavelength region (i).

Note that Born approximation is incorrectly applied in some modern literature on two-dimensional (2D) skyrmions in the limit $pR \to 0$. The scattering amplitude has a logarithmic non-analiticity in 2D in that limit \cite{LandauBook}, so the Born approximation is not applicable. . 

The eikonal approximation relies on two assumptions:  that semiclassical approximation is applicable and  that semiclassical trajectories may be approximated as straight lines. The semiclassical approximation is applicable when the momenta associated with the two bands are large, i.e. at $p R \gg 1$ and ${\left|(p_\Delta R)^2-(pR)^2\right| \gg 1}$. The straight-line approximation assumes that the momentum change due to the Lorentz force is much smaller than the magnitude of the initial momentum $p$. For a hopfion, that condition amounts to $pR \gg 1$ and coincides with the condition on semiclassics. As stated above, we focus on the case where only the bottom band is occupied, i.e. $p<p_\Delta$. The domain, where all these inequalities are satisfied, is shown as a shaded area (iii) in Fig.~\ref{fig:applicability}(b).

\subsection{Scattering amplitude in the eikonal approximation.} \label{sec:eikonal}
\begin{figure}
	(a)\includegraphics[width=0.95\linewidth]{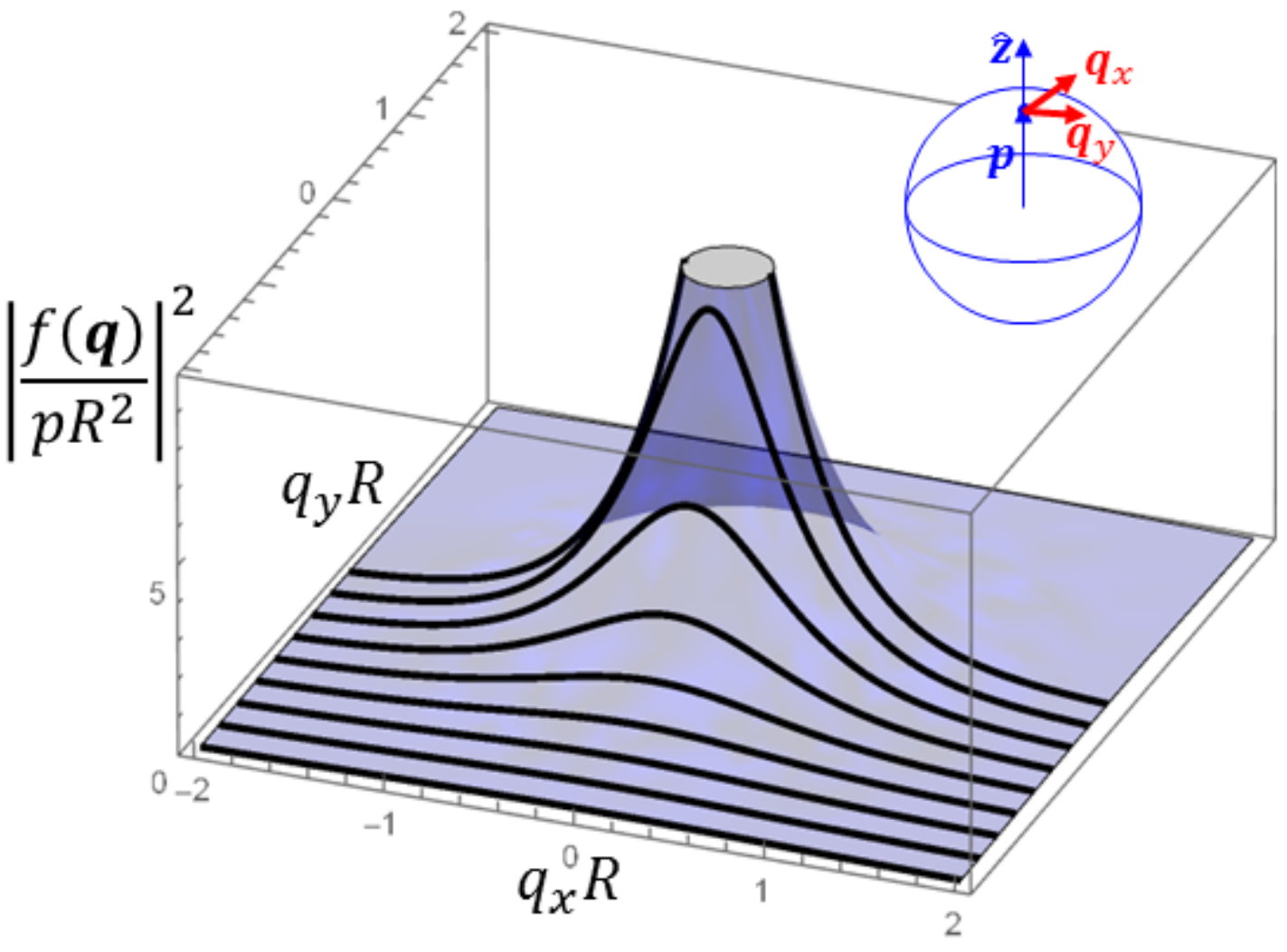} \\
	\vspace{0.5cm}
	(b)\includegraphics[width=0.95\linewidth]{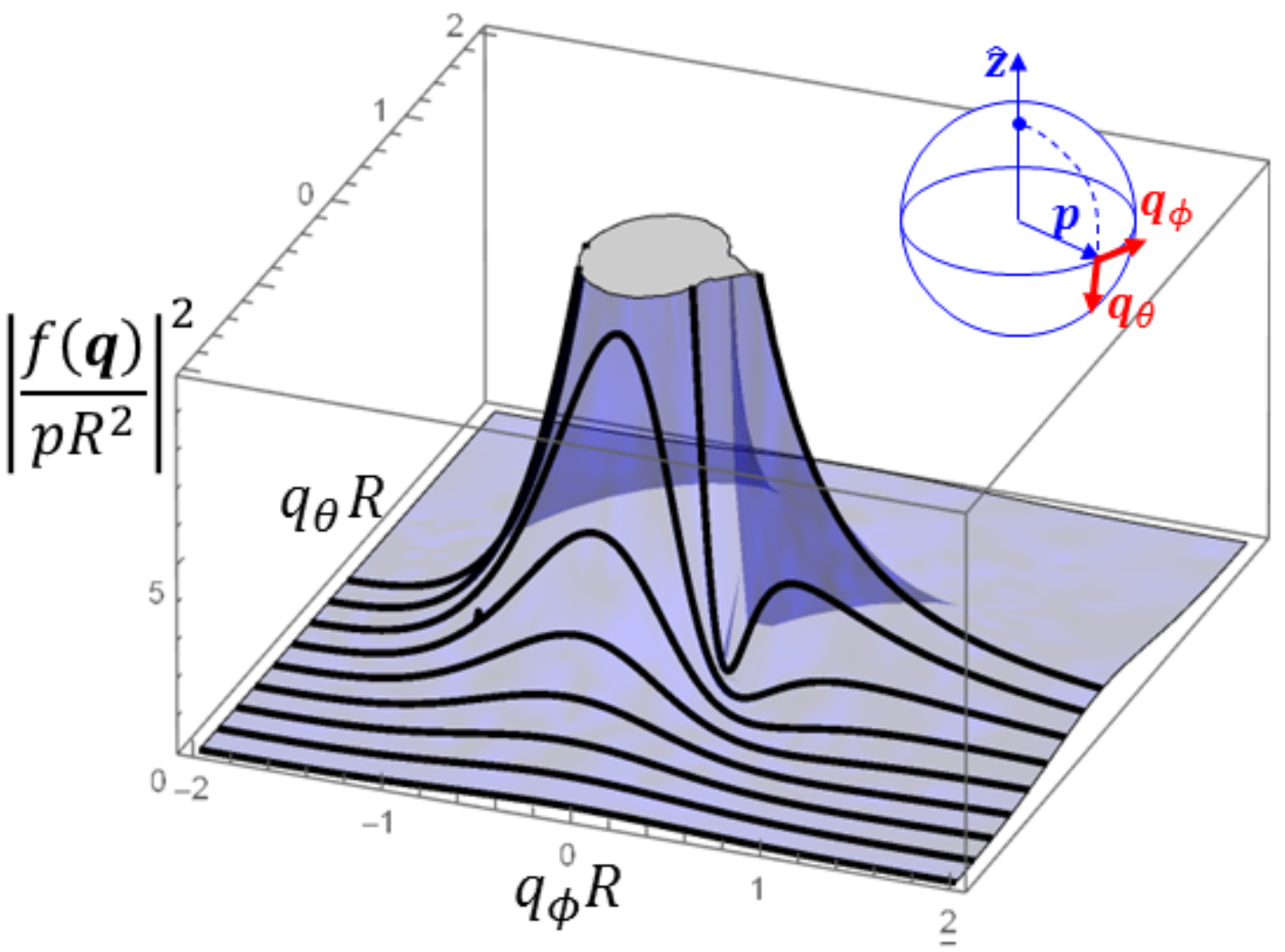}
	\caption{Normalized differential cross-section $|f_{\bm p}(\bm q)|^2$ vs momentum transfer $\bm q = \bm p'-\bm p$ evaluated using eikonal approximation. (a) In the case of the incoming electronic momentum parallel to the hopfion axis $\bm p\parallel \hat {\bm z}$, $|f_{\bm p}(\bm q)|^2$ is cylindrically symmetric. (b) Conversely, in the case of the incoming electronic momentum lying within the {\it hopfion plane} $\bm p\perp \hat {\bm z}$, the function $|f(\bm q)|^2$ is not cylindrically symmetric and contains a skew-scattering component (see discussion in the text below Eq.~(\ref{scat_cross}).
	The orientation of momentum $\bm p$ relative of the hopfion axes is illustrated in the top-right corners of the corresponding panels. } 
	\label{fig:eikonal}
\end{figure}
In this section, we use the eikonal approximation\cite{LandauBook} in order to evaluate the scattering amplitude in the domain (iii) shown in Fig.~\ref{fig:applicability}(b). In other words, we assume that $pR \gg 1$ and that only the bottom band, shown in Fig.~\ref{fig:applicability}(a), is occupied, i.e. $p < p_\Delta$. This limit is interesting because emergent fields $\bm A(\bm r)$ and $\bm B(\bm r)$ yield an appropriate  description of a scattering process. The original Hamiltonian~(\ref{hamiltonian}) reduces to $H = (\bm p- \bm A)^2/2m$, which allows to treat $\bm A(\bm r)$ as a conventional vector potential (electric charge $e$ is absorbed in the definition of $\bm A$). An additional merit of the eikonal approximation is that it is non-perturbative in potential $V$ and provides an easy way to capture the anisotropy of the hopfion texture. In this approximation, the electrons are slightly deflected from the original direction of propagation, i.e. the momentum transfer $\bm q = \bm p' - \bm p$ is small $q\sim 1/R \ll p$, where $\bm p$ and $\bm p'$ are the initial and finite momenta. That prompts us to denote the scattering amplitude as $f_{\bm p}(\bm q)$, where the subscript $\bm p$ emphasizes the initial momentum.

To simplify analytical calculation within this section, we use a hopfion profile~(\ref{hopfion})~with ${\eta(r) = \arccos[(R^2-r^2)(R^2+r^2)]}$ that produces a power-law decay $\delta S \sim R/r$ at $r \to \infty$. We substitute it in Eq.~(\ref{A}) and bring the latter equation to a concise form
\begin{align}
    \bm A(\bm r) = -4R^2\left[\bm r\,z\, \partial_R + \bm r\times \hat{\bm z}\right]\,\frac{1}{(r^2+R^2)^2}. \label{A_eikonal}
\end{align}
For clarity, here $z$ and $\hat {\bm z} = (0,0,1)$ denote the Cartesian coordinate and the unit vector aligned with the {\it hopfion axis}; $\partial_R \equiv \partial/\partial R$ is a partial derivative.

Scattering in the eikonal approximation may be understood as follows. Fast semiclassical electrons with momentum $\bm p$ are incident onto a magnetic texture and only slightly bend their trajectories. In the first approximation, their trajectories may be treated as straight lines. Let us label a given trajectory $T_{\bm p}(\bm \rho) = {\{\bm r = \bm \rho + \frac{\bm p}{m} t\}}$ by the momentum  $\bm p$  and the impact parameter $\bm \rho$. The latter is a vector residing in the plane perpendicular to $\bm p$, i.e. $\bm \rho \perp \bm p$; the origin $\bm \rho = 0$ is chosen at the center of the hopfion. Upon passing through the magnetic texture, electrons accumulate a (Berry) phase 
\begin{align}
    \delta_{\bm p}(\bm\rho) &= \int_{T_{\bm p}(\bm \rho)} d\bm r\cdot \bm A(\bm r) \nonumber \\
     & =\frac{2\pi R^2}{(R^2+\rho^2)^{3/2}}\,\left[ R\,(\hat{\bm p}\cdot\hat{\bm z})+\bm \rho\cdot(\hat{\bm p} \times \hat{\bm z})\right], \label{delta}
\end{align}
where the first line is the definition and the second is the result of substituting Eq.~(\ref{A_eikonal}).
Equation~(\ref{delta}) is practical because it yields a 2D map of a semiclassical phase for arbitrary direction of propagation $\hat{\bm p} = \bm p/p$. It is instructive to compare the phase~(\ref{delta}) with the emergent field profile shown in Fig.~\ref{fig:hopfion}(b). Let us consider electrons moving along the {\it hopfion axis} $\hat {\bm p}  = \hat{\bm z}$, so the second term in Eq.~(\ref{delta}) vanishes. Then the two equations $\delta_{\hat{\bm z}p}(\bm 0) = 2\pi$ and $\delta_{\hat{\bm z}p}(\hat{\bm x}\cdot\infty) = 0$ imply that the corresponding electron trajectories $T_{\hat{\bm z} p}(\bm 0)$ and $T_{\hat{\bm z} p}(\hat {\bm x}\cdot\infty)$ enclose an area with  magnetic flux quantum piercing through it. Indeed, the two lobes of magnetic field, shown in the left panel of Fig.~\ref{fig:hopfion}(b), carry quanta of magnetic flux of opposite signs, which produce phase accumulations of $2\pi$. Let us consider an incident electron traveling within the {\it hopfion plane}, e.g. $\hat{\bm p} = \hat{\bm x}$. Then only the second term in Eq.~(\ref{delta}) survives and renders $\delta_{\hat{\bm x} p}(\rho_y,\rho_z)$ an odd function of $\rho_y$. It is a consequence of the target-skyrmion-type profile of the field $B_z$ shown in the right panel of Fig.~\ref{fig:hopfion}(b). It is responsible for a skew-skattering component within the {\it hopfion plane} as will be shown below.  

Equation~(\ref{delta}) is used to evaluate the scattering amplitude\footnote{We note that the scattering amplitude~(\ref{amplitude_eikonal}) has a discontinuity at $\bm q=0$. It is a consequence of a long-range hopfion profile~(\ref{A_eikonal}) and disappears if a shorter-range profile is chosen. That discontinuity does not harm calculation of observables (e.g. optical theorem is satisfied).} in the eikonal approximation\cite{LandauBook}
\begin{align}
f_{\bm p}(\bm q) &= \frac{p}{2\pi i} \int{d^2\rho}\,\, e^{-i\bm q\cdot\bm \rho}\,\left[e^{i\delta_{{\bm p}}(\bm \rho)}-1\right]   \label{amplitude_eikonal}\\ 
&  =-ipR^2 \int_0^\infty ds\, s  \left\{\exp\left[\frac{2\pi i\, \hat{\bm  p}\cdot\hat{\bm z}}{(s^2+1)^{3/2}}\right] \right. \nonumber\\
& \left.\qquad \qquad\times J_0\left[\left|\frac{2\pi\,\hat{\bm  p}\times\hat{\bm z} }{(s^2+1)^{3/2}}-\bm qR\right|s\right] - J_0[qR\,s]\right\}, \nonumber
\end{align}
where, in the second equation, we integrated over the two-dimensional polar angle $\phi = \arctan(\rho_y/\rho_x)$, producing the Bessel function $J_0(z)$ of zeroth order, and reduced the integration to the dimensionless variable $s$. The value of equation~(\ref{amplitude_eikonal}) is that it provides a closed expression of the scattering amplitude for arbitrary incidence direction $\hat{\bm p}$. It is the central result of this section.

As a first application of Eq.~(\ref{amplitude_eikonal}), we evaluate the scattering cross-section 
\begin{align}
    &\sigma_{\bm p} \equiv \int \frac{d^2q}{p^2}|f_{\bm  p}(\bm q)|^2 =\label{scat_cross}
    \\ & 4\pi R^2\int_0^\infty ds\, s\,\left\{1-\cos\left[\frac{2\pi (\hat{\bm p}\cdot\hat{\bm z})}{(s^2+1)^{3/2}}\right]J_0\left[\frac{2\pi s \left|\hat{\bm p}\times\hat{\bm z}\right|}{(s^2+1)^{3/2}}\right]\right\}. \nonumber
\end{align}
We evaluate the integral numerically and find $\sigma_{\bm p \parallel \hat{\bm z}} \approx 7.17\, R^2$ and $\sigma_{\bm p \perp \hat{\bm z}} \approx 8.42\, R^2$ for an electron traveling along the hopfion axis $\hat {\bm z}$ and in the equatorial plane, respectively. It is a manifestation of the hopfion anisotropy. 

Let us inspect the differential cross-section $|f_{\bm p}(\bm q)|^2$ in the two opposite cases of electron traveling along the {\it hopfion axis}, i.e. $\bm p\parallel \hat{\bm z}$, and within the {\it hopfion plane}, i.e. $\bm p\perp \hat{\bm z}$. The case of intermediate angles may be understood as an interpolation between these two cases. We evaluate the integral~(\ref{amplitude_eikonal}) numerically and plot the results in Fig.~\ref{fig:eikonal}(a) and (b).  In the case $\bm p \parallel \hat{\bm z}$, the differential cross-section is cylindrically symmetric. It is a consequence of the cylindrical symmetry of hopfion configuration~(\ref{hopfion}) around the {\it hopfion axis} $\hat{\bm z}$. Conversely, for the electron traveling within the {\it hopfion plane} $\bm p \perp \hat{\bm z}$, the differential cross-section contains a skew-scattering  component. It is practical to expand the momentum-transfer vector $\bm q = (q_{\phi},q_{\theta})$ in the polar $q_\theta$ and azimuthal $q_\phi$ components [with respect to the {\it hopfion axis} $\hat{\bm z}$, see geometry in Fig.~\ref{fig:eikonal}(b)]. The scattering is  even in the polar $|f_{\bm p\perp \hat{\bm z}}(q_\phi,-q_\theta)|^2 = |f_{\bm p \perp \hat{\bm z}}(q_\phi,q_\theta)|^2$ and asymmetric in the azimuthal $|f_{\bm p \perp \hat{\bm z}}(-q_\phi,q_\theta)|^2 \neq |f_{\bm p \perp \hat{\bm z}}(q_\phi,q_\theta)|^2$ component. In other words, scattering has a skew-scattering component within the {\it hopfion plane} conventionally associated with the transverse Hall current. It is a consequence of the target-skyrmion-type profile of the magnetic field $B_z$ shown in the right panel of Fig.~\ref{fig:hopfion}(b). In order to quantify it, let us evaluate the corresponding  cross-section 
\begin{align}
    \sigma^{(1)}_{\bm p} &\equiv  \int \frac{d^2q}{p^2}\,\frac{q_{\phi}}{p}|f_{\bm  p}(\bm q)|^2   \nonumber  \\
    &=\int \frac{d^2\rho}{p}\,\partial_{\rho_{\phi}} \delta_{\bm p}(\bm \rho)   = 0  \label{sigma1}
\end{align}
where the first line is the definition, and the second line is the result of substituting Eq.~(\ref{amplitude_eikonal}). The superscript $1$ in Eq.~(\ref{sigma1}) indicates that the first power of momentum $q_\phi$ enters the integrand. The integral vanishes because the integrand is a full derivative of the continuous function $\delta_{\hat {\bm p}}(\bm \rho)$ that tends to a constant value $\delta_{\hat {\bm p}}(\bm \rho) \to 0$ at $\rho \to \infty$. It is consistent with a semiclassical observation\cite{Denisov2020} that the Hall current in the limit $pR \to \infty$ is proportional to the magnetic flux $\Phi$ piercing through the system. Since $\Phi \propto \langle B \rangle = 0$ for a hopfion (see Eq.~\ref{beq0}), the first-order skew-scattering cross-section $\sigma^{(1)}_{\bm p}$ vanishes in that limit. The first non-zero skew-scattering cross-section is of the third-order, i.e. $\sigma^{(3)}_{\bm p} \equiv \int \frac{d^2q}{p^2}\,\frac{q_{\phi}q_{\theta}^2}{p^3}|f_{\bm  p}(\bm q)|^2 \neq 0$.

Nevertheless, note that Eq.~(\ref{sigma1}) does not imply that the transverse Hall current vanishes for a hopfion. The precise equality $\sigma^{(1)}_{\bm p} = 0$ is a consequence of the eikonal approximation and the conditions of its applicability. As discussed in Sec.~\ref{sec:applicability}, those are that the semiclassical approximation is applicable and that trajectories may be approximated as straight lines. Departing from either of these conditions renders $\sigma^{(1)}_{\bm p} \neq 0$. In the following section, we demonstrate that a hopfion, indeed, has a skew-scattering cross-section $\sigma^{(1)}_{\bm p} \neq 0$ in the deeply quantum regime $pR \ll 1$. 

\subsection{Scattering amplitude in the Born approximation.} \label{sec:born}
In this Section, we examine scattering amplitude in the long-wavelength $pR \ll 1$ and weak-coupling ${p_\Delta R \ll 1}$ limit. To simplify the discussion, we focus on the case of a single Fermi surface, i.e. we further assume that the Fermi momentum is low $pR < p_\Delta R$. The combination of these conditions defines a shaded domain (i) in the space of parameters shown in Fig.~\ref{fig:applicability}(b). Then scattering may be analyzed using the Born series in $V = -\Delta\,\bm \sigma \cdot \delta\bm S(\bm r)$, see Eq.~(\ref{hamiltonian}). Below, we discuss results of Born series evaluated to second order. The details of the calculation are presented in Appendix~\ref{sec:born_appendix}.

To simplify calculation in this section, we rely on the hopfion profile~(\ref{hopfion}) with a Gaussian-type hopfion profile (see Appendix \ref{sec:born_appendix} for details). In the long-wavelength limit $pR \ll 1$ electrons do not resolve the fine spatial structure of the perturbation. As in electrodynamics, it is natural to analyze scattering in terms of moments. Therefore instead of the exact hopfion configuration~(\ref{hopfion}), we may use an approximate one 
\begin{align}
&\delta\bm S(\bm r)  =\left[\frac{a_1}{R}\left(\begin{array}{c}
         x  \\
         y \\
         0
\end{array}\right)  
- \frac{a_2}{R^2}\left(\begin{array}{c}
             -yz \\
              xz \\
              x^2+y^2
\end{array}\right)\right] e^{-r^2/R^2},  \label{hopfion_gauss_approx} 
\end{align}
where the dimensionless numerical coefficients ${a_1,a_2 \sim 1}$ play the role of moments. 

Anticipating the Born series, let us evaluate the diagonal $V_{\uparrow\uparrow }(\bm q) = -\Delta\, \delta S_z(\bm q)$ as well as the off-diagonal $V_{\downarrow\uparrow}(\bm q) = -\Delta\,[\delta S_x(\bm q) + i\,\delta S_y(\bm q)]$ and $V_{\uparrow\downarrow}(\bm q) = -\Delta\,[\delta S_x(\bm q) - i\,\delta S_y(\bm q)]$ matrix elements of the perturbation $V$ with respect to the plane-wave eigenstates of Hamiltonian~(\ref{hamiltonian}). To that end, we evaluate the Fourier transform of the Gaussian-type configuration~(\ref{hopfion_gauss_approx}) and obtain
\begin{equation}
\begin{aligned}
    V_{\uparrow\uparrow}(\bm q) &= -\pi^{3/2} \Delta\, R\, (\partial_{q_x}^2+\partial_{q_y}^2)\,e^{-q^2R^2/4} \\
    V_{\downarrow\uparrow}(\bm q) &= -i\pi^{3/2} \Delta\, R\, (\partial_{q_x}+i\partial_{q_y})(R\,a_1+a_2\,\partial_{q_z})\,e^{-q^2R^2/4}  \\
    V_{\uparrow\downarrow}(\bm q) &= -i\pi^{3/2} \Delta\, R\, (\partial_{q_x}-i\partial_{q_y})(R\,a_1-a_2\,\partial_{q_z})\,e^{-q^2R^2/4} 
\end{aligned} \label{matrix_elements}
\end{equation}
Here the derivatives $\partial_{q_x} \equiv \partial/\partial q_x$ and $\partial q_y \equiv \partial/\partial q_y$ originate from the terms $x$ and $y$ in the real space [see Eq.~(\ref{hopfion_gauss_approx})]. We commence with the first-order Born approximation, which is related to the Fourier transform $f^{(1)}(\bm n',\bm n) = -\frac{m}{2\pi}V_{\uparrow\uparrow}[p(\bm n'-\bm n)]$ (we use units $\hbar = 1$), where $\bm n = \bm p/p$ and $\bm n' = \bm p'/p'$ are the unit vectors in the direction of propagation of an incident and scattered electron. Within this section, it is practical to use a following notation for the scattering amplitude $f(\bm n',\bm n)$, where $\bm n$ and $\bm n'$ denote the unit vectors aligned with the direction of propagation of incident and scattered electrons. Expressing $\Delta = p_\Delta^2/4m$ and using Eq.~(\ref{matrix_elements}), we obtain
\begin{align}
    &f^{(1)}_{\uparrow\uparrow}(\bm n',\bm n) = \label{Born1}\\
    &\frac{\sqrt\pi a_2 p_\Delta^2 R^3}{8} \left\{-1 + (pR)^2\left[1-\bm n'\cdot\bm n -\frac 14(n_z'-n_z)^2 \right]\right\} \nonumber\\
    & \qquad\qquad\qquad\qquad\qquad\qquad\qquad\qquad\qquad\qquad+ \mathcal O(R^7),   \nonumber
\end{align}
where we also expanded in powers of $R$. Note that anisotropy of the hopfion configuration along the {\it hopfion axis} $\hat{\bm z}$ carries over to the anisotropy of scattering amplitude in that direction. In the second-order approximation, the scattering amplitude contains the no-spin-flip and spin-flip contributions
\begin{align}
    &f^{(2)}(\bm n',\bm n) = f^{(2)}_{\uparrow\uparrow\uparrow}(\bm n',\bm n) +  f^{(2)}_{\uparrow\downarrow\uparrow}(\bm n',\bm n) \label{Born2}\\
    &\quad=\frac{m^2}{\pi}\int \frac{d^3k}{(2\pi)^3}\,\frac{V_{\uparrow\uparrow}(\bm p' - \bm k)V_{\uparrow\uparrow}(\bm k - \bm p)}{k^2-p^2-i\delta} \nonumber \\
    &\qquad+\frac{m^2}{\pi}\int \frac{d^3k}{(2\pi)^3}\,\frac{V_{\uparrow\downarrow}(\bm p' - \bm k)V_{\downarrow\uparrow}(\bm k - \bm p)}{k^2+p_\Delta^2-p^2-i\delta}. \nonumber 
\end{align}
The sign of the infinitesimal imaginary part in the denominators accounts for causality in the scattering theory~\cite{LandauBook}. It may be dropped in the spin-flip term $f^{(2)}_{\uparrow\downarrow\uparrow}(\bm p',\bm p)$ due to the assumed condition $p_\Delta > p$. We evaluate the integrals above in Appendix~\ref{sec:born_appendix} and expand in powers of $R$: 
\begin{widetext}
\begin{align}
f^{(2)}_{\uparrow\uparrow\uparrow}(\bm n',\bm n) &= \frac{\sqrt\pi \,p_\Delta^4\, R^5}{4!\,8 \sqrt 2} \left\{c_1 +3\,i \,(pR)\sqrt{2\pi} \right. \label{Born2_uuu}\\
&\left.+(pR)^2\left[-c_2+c_3\bm n'\cdot\bm n - c_4 n_z'n_z+ c_5(n_z'^2+n_z^2) \right]\right\} + O(R^8), \nonumber \\
f^{(2)}_{\uparrow\downarrow\uparrow}(\bm n',\bm n) &= \frac{\sqrt\pi p_\Delta^4 R^5}{4!\,8 \sqrt 2}\left\{ c_6 + c_7\, (pR)\, (n_z'+n_z)-c_8 (p_\Delta R)^2\right.   \label{Born2_udu} \\
& \left.+(pR)^2\left[-c_9+c_{10}\, \bm n'\cdot \bm n + c_{11}\, n_z'n_z+ c_{12}\,(n_z'^2+n_z^2)+i\, c_{13}\,(\bm n'\times \bm n)_z\right]\right\}+ O(R^8) \nonumber 
\end{align}
\end{widetext}
The second-order Born correction~(\ref{Born2_uuu})-(\ref{Born2_udu}) has a rich angular structure. A few comments are in order. (i) The dimensionless
coefficients $c_1$ - $c_{13}$ are numbers of order 1 and depend on the details of the hopfion structure at short-range scale. Their specific values are listed in Appendix~\ref{sec:born_appendix}. (ii) The imaginary part of $f_{\uparrow\uparrow\uparrow}^{(2)}$ is universal (i.e. independent of short-scale geometry of the hopfion) and originates from the on-shell processes in the denominator in Eq.~(\ref{Born1}). It satisfies the optical theorem and together with the first-order Born result~(\ref{Born1}) serves as an additional verification of Eqs.~(\ref{Born2_uuu})-(\ref{Born2_udu}). (iii) The scattering amplitude is anisotropic due to the anisotropy of the hopfion profile. (iv) The lowest~order $\propto R^{5}$ terms are scalars, which produce s-wave scattering. The first term with non-trivial angular dependence $c_7(n_z'+n_z)$ appears in the order $\propto R^6$ in $f_{\uparrow\downarrow\uparrow}^{(2)}$. It is odd both under inversion and time-reversal transformations. We interpret it as a scattering due to the {\it toroidal moment} of the hopfion (see the following Section). (v) Observe that the skew-scattering term $\propto i c_{13}(\bm n'\times \bm n)_z$ is generated in $f_{\uparrow\downarrow\uparrow}$. Its interference with the imaginary term in Eq.~(\ref{Born2_uuu}) produces a skew-scattering term in the differential cross-section $|f(\bm n',\bm n)|^2$, which results in the non-zero Hall effect. 
\subsection{Possible transport signatures.} \label{sec:transport}
\subsubsection{Toroidal scattering.} \label{sec:toroidal}
Below, we discuss the origin and possible experimental signature of the {\it toroidal} term $\propto c_7 (n_z + n_z')$ appearing Eq.~(\ref{Born2_udu}). In order to motivate its naming, let us evaluate the scattering amplitude in the first-order Born approximation in the vector-potential part $(\bm p\cdot\bm A+\bm A\cdot\bm p)/2m$ of the Hamiltonian $H = (\bm p-\bm A)^2/2m$. We get \cite{GalitskiBook},\footnote{The applicability of the Born series expansion in $\bm p\cdot \bm A$ was not investigated.} 
\begin{align}
    f(\bm p',\bm p) = \frac{1}{4\pi} (\bm p'+\bm p)\cdot\bm A_{\bm p'-\bm p}, \label{scat_ampl_pA}
\end{align}
where we used the units with $\hbar = 1$. 
Equation~(\ref{scat_ampl_pA}) implies that the expansion of the scattering amplitude $f(\bm p',\bm p)$ in small $\bm p$ may be obtained from the corresponding expansion of the vector potential $A_{\bm p-\bm p'}$ Fourier transform. Let us show that the zeroth order term in the latter expansion is related to the toroidal moment $\bm L$ given by Eq.~(\ref{toroidal}). To that end, let us express the toroidal moment (\ref{toroidal}) in terms of the vector potential as follows $\bm L = \frac 12\int d^3r\, [\bm r \times \bm \nabla \times \bm A(\bm r)]$. Integrating that integral by parts, we may flip the gradient operator $\bm \nabla$ to act onto $\bm r$, which reduces the integral to $\bm L = \int d^3r \bm A(\bm r) \equiv \int d^3r \bm A(\bm r) e^{-i\bm 0\cdot \bm r} = \bm A_{\bm q=\bm 0}$. In other words, the leading order term in the $\bm q$ expansion of the vector potential $\bm A_{\bm q}$ is the toroidal moment $\bm L$, i.e. $\bm A_{\bm q} \approx \bm L + \mathcal O(\bm q)$. Substituting the latter equation in Eq.~(\ref{scat_ampl_pA}), one may find the small momentum expansion of the scattering amplitude as
\begin{align}
   f(\bm p',\bm p) &= \frac{1}{4\pi} (p'_z+ p_z)L ,\label{scat_ampl_pA_1}
\end{align}
where we used that a hopfion toroidal moment $\bm L$ is  aligned with the {\it hopfion axis} $\hat{\bm z}$. 
The provided first-order Born calculation in $\bm p\cdot \bm A$ is distinct from that in Sec.~\ref{sec:born}, which produced Eqs.~(\ref{Born2_uuu})-(\ref{Born2_udu}). The former assumes that we are deep in the adiabatic region $p_\Delta R \gg 1$, whereas the latter assumes $p_\Delta R \ll 1$ (see the applicability diagram in Fig.~\ref{fig:applicability}(b)). However, comparison of Eq.~(\ref{scat_ampl_pA_1}) with Eq.~(\ref{Born2_udu})  offers an intuition that $\propto c_7 (n_z + n_z')$ is generated by the toroidal moment $\bm L$.  Similar to the toroidal vector $\bm L$, the term $\propto c_7 (n_z + n_z')$ is odd under time-reversal and inversion operations. When the total differential cross-section is evaluated $|f(\bm n',\bm n)|^2 = |f^{(1)}_{\uparrow\uparrow}(\bm n',\bm n)+f^{(2)}_{\uparrow\uparrow\uparrow}(\bm n',\bm n)+f^{(2)}_{\uparrow\downarrow\uparrow}(\bm n',\bm n)+\cdots|^2$, the term due to the toroidal scattering  $\propto c_7 (n_z + n_z')$ interferes with s-wave terms [such as e.g. $\propto c_1$ and $\propto c_6$ in Eqs.~(\ref{Born2_uuu})-(\ref{Born2_udu})] and renders the differential cross-section $|f(\bm n',\bm n)|^2$ a non-reciprocal \cite{Nagaosa2018,Mannhart2018} function of $\bm n$ and $\bm n'$. Specifically, the differential cross-section for electrons propagating in $\hat{\bm z}$ and $-\hat{\bm z}$ directions is distinct. It is conceivable\cite{Nagaosa2018,Mannhart2018} that a device containing a hopfion could exhibit a diode-type behavior along the {\it hopfion axis} $\hat{\bm z}$. In other words, the I-V curve could be asymmetrical in the applied bias voltage $V_z$, i.e. $I_z(V_z) \approx G_0 V_z+G_1V_z^2+\mathcal O(V_z^3)$. Here, the second-order conductance $G_1$ is induced by the toroidal moment $\bm L$.
 
\subsubsection{Skew-scattering term.} \label{sec:skew-scattering}
Observe that Eq.~(\ref{Born2_udu}) contains a skew scattering term $i\, c_{13}\,(\bm n'\times \bm n)_z$. When the total differential-cross-section is evaluated $|f(\bm n',\bm n)|^2 = |f^{(1)}_{\uparrow\uparrow}(\bm n',\bm n)+f^{(2)}_{\uparrow\uparrow\uparrow}(\bm n',\bm n)+f^{(2)}_{\uparrow\downarrow\uparrow}(\bm n',\bm n)+\cdots|^2$, the skew-scattering term $i\, c_{13}\,(\bm n'\times \bm n)_z$ interferes with an imaginary s-wave scattering term~$3\,i \,(pR)\sqrt{2\pi}$ in (\ref{Born2_uuu}). That also produces a skew-scattering term $\propto (\bm n'\times\bm n)_z$ in the differential cross-section $|f(\bm n',\bm n)|^2$ responsible for the Hall effect within a {\it hopfion plane}. We note that the Hall conductance in the hopfion plane was indeed observed in a numerical Landauer-Buttiker calculation of a hopfion in a mesoscopic setting \cite{Mertig2020}. However, it was interpreted as an artifact due to the discretization of the hopfion on a lattice. In contrast, we argue that the Hall effect is an intrinsic property of the hopfion, which arises due to the target-skyrmion-type profile of the magnetic field, see right panel in Fig.~\ref{fig:hopfion}(b).

\section{Conclusion} \label{sec:conclusion}
A hopfion is a topological configuration~(\ref{hopfion}) of the vector $\bm S(\bm r)$. It has been long sought after in magnetic \cite{wilczek1983,Fadeev1997,Zang2020} and other systems \cite{Babaev2002,hopfion_ferroelectric}. There is a claim of the first observation in experiment \cite{hopfion_experiment}. Motivated by an analogy with the Aharonov-Bohm scattering~\cite{AB_paper} in 2D, we study a hitherto unexplored question of itinerant electrons scattering off a topological object (hopfion) in 3D. Because a hopfion is anisotropic, the conventional method of decomposing in partial waves with distinct angular harmonics is not applicable. So we resort to eikonal and  (second-order)  Born approximations applicable in the opposite limits $pR \gg 1$ and $pR \ll 1$. Both approaches show that a scattering amplitude has a rich angular structure induced by the hopfion anisotropy. We find that, although the average magnetic field~(\ref{beq0}) vanishes, the differential cross-section does have a skew-scattering component within the {\it hopfion plane} z = 0. It is associated with a target-skyrmion-type structure of the emergent magnetic field induced by the hopfion, shown in the right panel of Fig.~\ref{fig:hopfion}(b). It leads to the non-vanishing Hall effect within the hopfion plane (see Sec.~\ref{sec:skew-scattering}). In the $pR \ll 1$ limit, we find a term due to the hopfion toroidal moment $\bm L$. It is odd both under the time-reversal and inversion operations. It may produce a non-reciprocal response in a device containing a hopfion, i.e. the I-V curve $I_z(V_z) \approx G_0 V_z+G_1V_z^2+\mathcal O(V_z^3)$ contains the second order term in the bias voltage $V_z$ applied along the hopfion axis (see Sec.~\ref{sec:toroidal}). The developed methods are applicable to other 3D magnetic structures, which will be explored experimentally \cite{Donnelly2017} in the near future. 

This work was supported by the U.S. Department of Energy (DOE), Office of Science, Basic Energy Sciences (BES) under Award No. DE-SC0020221.

\bibliography{biblio}

\appendix

\section{Details of calculation of the second Born approximation~(\ref{Born2}).} \label{sec:born_appendix}
\begin{figure}
    \includegraphics[width=0.95\linewidth]{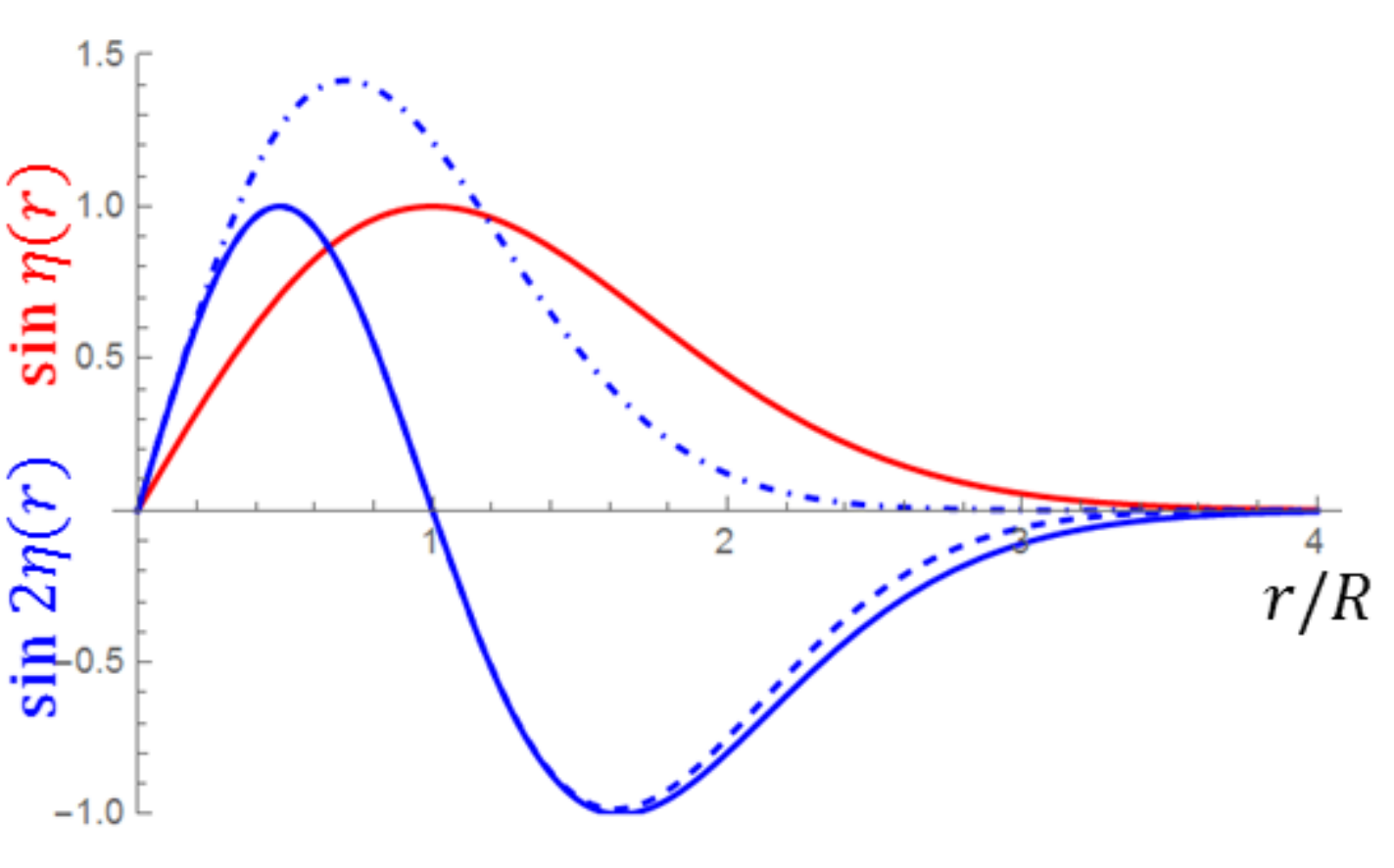} 
	\caption{Dependence of $\sin \eta(r)$ and $\sin 2\eta(r)$, which specifies a hopfion profile~(\ref{hopfion_b}). Solid lines correspond to the exact Eq.~(\ref{sineta}), whereas dashed and dashed-dotted lines correspond to different approximations of $\cos \eta(r)$ [see main text around Eq.~(\ref{cos_approx})].}  
	\label{fig:sineta}
\end{figure}

In this section, we provide the details of evaluating the second-order Born approximation in perturbation $V$ (see Eq.~(\ref{hamiltonian})) for a hopfion configuration~(\ref{hopfion}).   

(i) The strategy is to reduce hopfion spatial configuration~(\ref{hopfion}) to a gaussian-type profile, for which integrals~(\ref{Born2}) may be evaluated carefully. We write the hopfion configuration as
\begin{align}
&\delta\bm S(\bm r)  =\left(\begin{array}{c}
         x  \\
         y \\
         0
\end{array}\right)\frac{\sin 2\eta(r)}{r} - \left(\begin{array}{c}
             -yz \\
              xz \\
              x^2+y^2
\end{array}\right) \frac{2\sin^2\eta(r)}{r^2}.  \label{hopfion_b} 
\end{align}
Let us choose a specific form for the trigonemetric functions appearing in Eq.~(\ref{hopfion_b})
\begin{equation}
\begin{aligned}
    &\sin\eta(r) = \frac{r}{R} \, e^{\frac{1}{2}\left(1-\frac{r^2}{R^2}\right)}, \\
    &\cos\eta(r) = {\rm sign}(R-r)\sqrt{1- \left(\frac rR\right)^2e^{\left(1-\frac{r^2}{R^2}\right)}}. 
\end{aligned} \label{sineta}
\end{equation}
The two functions $\sin \eta(r)$ and $\sin 2\eta(r)$, which appear in Eq.~(\ref{hopfion_b}), are plotted in Fig.~\ref{fig:sineta}. As intended, they correspond to a monotonic $\eta (\bm r)$ ranging from $0$ to $\pi$ as $r$ goes from $0$ to $\infty$. Observe that $\sin \eta$ is a product of $r$ and a Gaussian function $e^{-r^2/2R^2}$, which renders it convenient for integration (performed below). In contrast, $\cos \eta(r)$, which ranges from $1$ to $-1$ as $r$ goes from $0$ to $\infty$, is not easily reduced to a Gaussian.  Nevertheless, observe that $\cos\eta(r)$ enters Eq.~(\ref{hopfion_b}) via $\sin 2\eta(r) = 2\sin\eta(r) \cos\eta(r)$. Due to that and to the fact that $\sin\eta(r)$ is exponentially-localized (see Fig.~\ref{fig:sineta}), we do not need a uniform approximation of $\cos\eta(r)$. We may approximate $\cos \eta(r)$ as a product of a Gaussian and polynomial of $r^2$
 \begin{align} 
    \cos \eta (r) \approx e^{-\frac{r^2}{2R^2}} \sum_{n = 0}^{N} c_n\, \left(\frac{r}{R}\right)^{2n}. \label{cos_approx}  
 \end{align}
For example, setting $N=4$ and evaluating coefficients $c_n$ produces a very good approximation for $\sin 2\eta(r)$ shown with a dashed line in Fig.~\ref{fig:sineta}. Further increase of $N$ produces an approximation for $\sin 2\eta$ indistinguishable from the exact result. To  simplify analytical calculations, we truncate the polynomial in Eq.~(\ref{cos_approx}) to $N=0$ and set the only coefficient $c_0 = 1$. It yields $\sin 2\eta(r)$ plotted with a dash-dotted line in Fig.~\ref{fig:sineta}. A significant disparity between that approximation and the exact dependence (solid dashed line) is not essential since we are interested in evaluating the long-wavelength behavior $pR \ll 1$. To conclude this paragraph, setting $\cos \eta (r) = e^{-r^2/2R^2}$ and using $\sin\eta(r)$ from Eq.~(\ref{sineta}) allows to write Eq.~(\ref{hopfion_b}) as
\begin{align}
&\delta\bm S(\bm r)  =\left[\frac{a_1}{R}\left(\begin{array}{c}
         x  \\
         y \\
         0
\end{array}\right)  
- \frac{a_2}{R^2}\left(\begin{array}{c}
             -yz \\
              xz \\
              x^2+y^2
\end{array}\right)\right] e^{-r^2/R^2},  \label{hopfion_c} 
\end{align}
where the dimensionless coefficients $a_1 = 2\sqrt{e}$ and $a_2 = 2e$ are introduced to keep track the contribution of the distinct terms in the calculations below. Observe that equation~(\ref{hopfion_c}) is a product of a Gaussian and simple polynomials of coordinates $(x,y,z)$. As such it is amenable for the analytical calculation performed below.

(ii) Anticipating the Born approximation, we evaluate the Fourier transform of Eq.~(\ref{hopfion_c})
\begin{align}
    &\delta \bm S(\bm q) \label{hopfion_d}\\
    &= \pi^{3/2}R^3 \left[\frac{a_1}{R}\left(\begin{array}{c}
         i\partial_{q_{x}}  \\
         i\partial_{q_{y}} \\
         0
\end{array}\right)  
+ \frac{a_2}{R^2}\left(\begin{array}{c}
             -\partial_{q_y}\partial_{q_z} \\
              \partial_{q_x}\partial_{q_z} \\
              \partial_{q_x}^2+\partial_{q_z}^2 
\end{array}\right)\right] e^{-\frac{(qR)^2}{4}} \nonumber
\end{align}
where we retain the momentum derivatives~$\bm \partial_q = \left(\frac{\partial}{\partial q_x},\frac{\partial}{\partial q_y},\frac{\partial}{\partial q_z}\right)$. Using Eq.~(\ref{hopfion_d}), we may also explicitly write the matrix elements of the perturbation
\begin{align}
    V_{\uparrow\uparrow}(\bm q) &\equiv -\Delta \delta S_z(\bm q) \label{vuu}\\
    &= -\pi^{3/2} \Delta\, R\, (\partial_{q_x}^2+\partial_{q_y}^2)\,e^{-q^2R^2/4} \nonumber \\
    V_{\downarrow\uparrow}(\bm q) &\equiv -\Delta \left[\delta S_x(\bm q)+iS_y(\bm q)\right] \label{vdu}\\
    &= -i\pi^{3/2} \Delta\, R\, (\partial_{q_x}+i\partial_{q_y})(R\,a_1+a_2\,\partial_{q_z})\,e^{-q^2R^2/4} \nonumber \\
    V_{\uparrow\downarrow}(\bm q) &\equiv -\Delta \left[\delta S_x(\bm q)-iS_y(\bm q)\right] \label{vud}\\
    &= -i\pi^{3/2} \Delta\, R\, (\partial_{q_x}-i\partial_{q_y})(R\,a_1-a_2\,\partial_{q_z})\,e^{-q^2R^2/4} \nonumber
\end{align}

(iii) {\it First-order Born approximation.} The scattering amplitude in the first-order Born approximation may be evaluated (in units $\hbar =1$)
\begin{align*}
    f^{(1)}_{\uparrow\uparrow} & = -\frac{m}{2\pi}V_{\uparrow\uparrow}(\bm q) \\ 
    & =\frac{\sqrt \pi\, a_2\,p_\Delta^2\,R}{8}\left(\partial_{q_x}^2+\partial_{q_y}^2\right) e^{-q^2R^2/4}
\end{align*}
where $\bm q = \bm p'-\bm p$ is the momentum transfer; $\bm p$ and $\bm p'$ are the momenta of the initial and finite state. Further, we denote $\bm p' = p \, \bm n'$ and $\bm p = p\, \bm n$ and expand the equation above in powers of $R$
\begin{align}
    &f^{(1)}_{\uparrow\uparrow}(\bm n',\bm n) = \\
    &\frac{\sqrt\pi a_2 p_\Delta^2 R^3}{8} \left\{-1 + (pR)^2\left[1-\bm n'\cdot\bm n -\frac 14(n_z'-n_z)^2 \right]\right\} \nonumber\\
    & \qquad\qquad\qquad\qquad\qquad\qquad\qquad\qquad\qquad\qquad+ \mathcal O(R^7).   \nonumber
\end{align}

(iv) {\it Second-order Born approximation: the no-spin-flip contribution}. Now, let us evaluate the no-spin-flip part of the second-order Born approximation 
\begin{align}
    f^{(2)}_{\uparrow\uparrow\uparrow}(\bm p',\bm p) &= \frac{m^2}{\pi} \int \frac{d^3k}{(2\pi)^3}\,\frac{V_{\uparrow\uparrow}(\bm p'-\bm k)V_{\uparrow\uparrow}(\bm k-\bm p)}{k^2-p_E^2-i\delta} \nonumber \\
    &= \frac{\pi^2a_2^2p_\Delta^4R^2}{16} (\partial_{p_x'}^2+\partial_{p_y'}^2)(\partial_{p_x}^2+\partial_{p_y}^2) I(\bm p',\bm p), \label{f_uuu_interm} \\
    & {\rm where\,\,} I(\bm p',\bm p) = \int \frac{d^3k}{(2\pi)^3} \frac{e^{-[(\bm p'-\bm k)^2+(\bm k-\bm p)^2]R^2/4}}{k^2-p_E^2-i\delta}. \nonumber
\end{align}
Here, we substitute the matrix element~(\ref{vuu}) and pulled the derivatives over the ``external'' momenta outside the integral sign. In the denominator of the integrand, we used a distinct notation $p_E = \sqrt{2mE}$ to distinguish it from the variables $\bm p$ and $\bm p'$, over which the derivatives are taken. We set $p_E \to p$ at the end of the calculation. The integral~$I(\bm p',\bm p)$ is evaluated in Appendix~\ref{sec:integral_appendix}. In principle, Eq.~(\ref{f_uuu_interm}) contains complete information about the second-order scattering amplitude $f^{(2)}_{\uparrow\uparrow\uparrow}$. However, we are interested in the small-$R$ expansion 
\begin{align}
&f^{(2)}_{\uparrow\uparrow\uparrow}(\bm n',\bm n) = \frac{\sqrt\pi \,p_\Delta^4\, R^5}{192\sqrt 2} \left\{c_1 +3\,i \,(pR)\sqrt{2\pi} \right. \nonumber\\
&\left.+(pR)^2\left[-c_2+c_3\bm n'\cdot\bm n - c_4 n_z'n_z+ c_5(n_z'^2+n_z^2) \right]\right\}  \nonumber \\
&\qquad\qquad\qquad\qquad\qquad +\mathcal O(R^8),\label{f_uuu}
\end{align}
where $c_1 = 23a_2^2/5$, $c_2 = 1677a_2^2/140$, $c_3 = 157a_2^2/140$, $c_4 =-114a_2^2/140$ and $c_5 = 153a_2^2/140$ are the numerical coefficients. Observe that, to the lowest order in $R$, the imaginary part of the second-order amplitude satisfies the optical theorem, i.e. $\sigma_{\propto R^6} = \frac{4\pi}{p} {\rm Im} f^{(2)}_{\uparrow\uparrow\uparrow} = 4\pi \left|{f^{(1)}_{\uparrow\uparrow}}_{\propto R^3}\right|^2$. It serves as an independent verification of the numerical coefficients.

(v) Similarly, we may evaluate the {\it spin-flip cotribution} to the second-order scattering amplitude
\begin{align}
    &f^{(2)}_{\uparrow\downarrow\uparrow}(\bm p',\bm p) = \frac{m^2}{\pi} \int \frac{d^3k}{(2\pi)^3}\,\frac{V_{\uparrow\downarrow}(\bm p'-\bm k)V_{\downarrow\uparrow}(\bm k-\bm p)}{k^2-p_E^2-i\delta} \nonumber \\
    & = \frac{\pi^2a_2^2p_\Delta^4R^2}{16} (\partial_{p_x'}-i\partial_{p_y'})(Ra_1-a_2\partial_{p_z'}) \nonumber\\ & \qquad\qquad\qquad\qquad\times(\partial_{p_x}+i\partial_{p_y})(Ra_1-a_2\partial_{p_z})\tilde I(\bm p',\bm p), \nonumber \\
    & \tilde I(\bm p',\bm p) = \int \frac{d^3k}{(2\pi)^3} \frac{e^{-[(\bm p'-\bm k)^2+(\bm k-\bm p)^2]R^2/4}}{k^2+ p_\Delta^2-p_E^2} \nonumber,
\end{align}
where we substitute the matrix-elements~(\ref{vdu})-(\ref{vud}), pulled the derivatives outside of the integral $\tilde I$. The latter integral may be obtained from the integral $I$, given by Eq.~(\ref{Ipp}), by the substitution 
\begin{align*}
 p_E \to i\kappa, \quad \kappa = \sqrt{p_\Delta^2-p_E^2}.
\end{align*}
So, we may obtain an expansion of the amplitude in powers of $R$
\begin{widetext}
\begin{align}
    f^{(2)}_{\uparrow\downarrow\uparrow}(\bm n',\bm n) &= \frac{\sqrt\pi p_\Delta^4 R^5}{192\sqrt 2}\left\{ c_6 + c_7\, (pR)\, (n_z'+n_z)-c_8(\kappa R)^2\right.   \nonumber \\
    & +(pR)^2\left[-c_9+c_{10}\, \bm n'\cdot \bm n + c_{11}\, n_z'n_z+ c_{12}\,(n_z'^2+n_z^2)+ic_{13}\,(\bm n'\times \bm n)_z\right] \nonumber
\end{align}
\end{widetext}
where $c_6 = a_1^2 + 3a_2^2/20$, $c_7= a_1a_5/5$, $c_8 = a_1^2 + a_2^2/20$, $c_9 = 7a_1^2/10+27a_2^2/280$, $c_{10} = 13a_1^2/10 + 29 a_2^2/280$, $c_{12} = 7a_1^2/40-27a_2^2/1120$ and $c_{13} = a_1^2/28 + a_2^2/20$ are numerical coefficients of order 1.
\section{Evaluation of the integral } \label{sec:integral_appendix}
In this Section, we evaluate the integral
\begin{align}
    I(\bm p',\bm p) = \int \frac{d^3k}{(2\pi)^3} \frac{e^{-[(\bm p'-\bm k)^2+(\bm k-\bm p)^2]R^2/4}}{k^2-p_E^2-i\delta}.
\end{align}
It may arise in other applications involving second-order Born approximation for a Gaussian-type potential. So, it is worth to provide the details of integration.

(i) We introduce auxiliary momentum variables $\bm Q = \frac{\bm p'-\bm p}{2}$ and $\bm l = \frac{\bm p'+\bm p}{2}$ and integrate $I$ over the angles 
\begin{align*}
    I = \frac{e^{-Q^2R^2/2}}{(2\pi)^2lR^2} \int_{-\infty}^\infty dk\, \frac{k\,e^{-(k-l)^2R^2/2}}{k^2-p_E^2-i\delta}, 
\end{align*}
where we also extended the limits of $k$-integration to $(-\infty, 0)$. 

(ii) The rational function of $k$ in the integrand may be split as follows 
\begin{align*}
    &\frac{k}{k^2-p_E^2-i\delta} = \frac 12\left[\frac{1}{k-p_E-i\delta}+\frac{1}{k+p_E+i\delta}\right] \\
    &= \frac 12\left[\pi i\,\delta(k-p_E)-\pi i\,\delta(k+p_E)+\frac{1}{k-p_E}+\frac{1}{k+p_E}\right],
\end{align*} 
where we applied the Sokhotski formula in the second line to split the imaginary and real parts. The integration over the former is then evaluated exactly, whereas the latter produce principal value integrals 
\begin{align*}
    I &= \frac{e^{-Q^2R^2/2}}{2(2\pi)^2lR^2} \left\{2\pi i\, \sinh(lp_ER^2) e^{-(l^2+p_E^2)R^2/2}\right. \\
    &\left.+ {\rm v.p.}\int_{-\infty}^\infty dk\,\frac{e^{-(k-l)^2R^2/2}}{k-p_E} + {\rm v.p.}\int_{-\infty}^\infty dk\,\frac{e^{-(k-l)^2R^2/2}}{k+p_E} \right\} \\
    &= \frac{e^{-Q^2R^2/2}}{2(2\pi)^2lR^2} \left\{2\pi i\, \sinh(lp_ER^2) e^{-(l^2+p_E^2)R^2/2}\right. \\
    &\qquad\quad+ {\rm v.p.}\int_{-\infty}^\infty dk\,\frac{e^{-(k-(l-p_E))^2R^2/2}}{k} \\ 
    &\qquad\qquad\left.+ {\rm v.p.}\int_{-\infty}^\infty dk\,\frac{e^{-(k-(l+p_E))^2R^2/2}}{k} \right\} \\
    &= \frac{e^{-Q^2R^2/2}}{2(2\pi)^2lR^2} \left\{2\pi i\, \sinh(lp_ER^2) e^{-(l^2+p_E^2)R^2/2}\right. \\
    & \qquad\qquad\left. +J[(l-p_E)R] + J[(l+pE)R] \right\}
\end{align*}
where we shifted the integration variable in the penultimate and  defined the function 
\begin{align}
J(t) \equiv {\rm v.p.}\int_{-\infty}^{+\infty} dk\, \frac{e^{-(k-t)^2/2}}{k}.
\end{align}

(iii) Note that $J(t)$ is not an elementary function. Let us evaluate its Taylor expansion in $t$. We introduce an auxiliary parameter $\lambda$ and upgrade to a new function 
\begin{align*}
    \tilde J(t,\lambda) \equiv  {\rm v.p.}\int_{-\infty}^{+\infty} dk \frac{e^{-(k^2-2tk\lambda+t^2)/2}}{k}
\end{align*}
such that its derivative over $\lambda$ may be easily evaluated by taking the Gaussian integral
\begin{align*}
    \frac{d\tilde J(t,\lambda)}{d\lambda}  &= {\rm v.p.}\int_{-\infty}^{+\infty} dk \,t\,e^{-(k^2-2tk\lambda+t^2)/2} \\
    &= \sqrt{2\pi}\, t\,e^{-t^2(1-\lambda^2)/2}.
\end{align*}
In addition, noting that $\tilde J(t,1) = J(t)$ and $\tilde J(t,0) = 0$ allows us to obtain
\begin{align}
    J(t) &= \int_0^1 d\lambda\,\frac{d\tilde J(t,\lambda)}{d\lambda} \nonumber \\
    &= \sqrt{2\pi}\,t\int_0^1 d\lambda\,e^{-t^2(1-\lambda^2)/2} \nonumber \\
    &=\sum_{n = 0}^{\infty} a_n\, t^{2n+1}, \,\, a_n = \frac{\sqrt{2\pi}}{2^n}\sum_{m=0}^n\,\frac{(-1)^{n+m}}{m!\,(n+m)!\,(2m+1)}, 
\end{align}
where, in the second line, we expand the exponent in the Taylor series and integrate it term-by-term, which produces the expansion in the third line.

(iv) This concludes the evaluation of the intergal $I(\bm p',\bm p)$. Let us write it explicitly in the variables $\bm p$ and $\bm p'$
\begin{align}
    &I(\bm p,\bm p') = \frac{\exp\left[-\frac{(\bm p' - \bm p)^2R^2}{8}\right]}{(2\pi R)^2|\bm p+\bm p'|} \label{Ipp} \\
    &\times\left\{2\pi i\, \sinh\left[\frac{|\bm p'+\bm p|p_ER^2}{2}\right]\exp\left[\frac{(\bm p'+\bm p)^2R^2+4p_E^2R^2}{8}\right]\right.\nonumber \\
    & \left.+J\left[\frac{1}{2}|\bm p'+\bm p|R - p_ER\right]+J\left[\frac{1}{2}|\bm p'+\bm p|R + p_ER\right]\right\}. \nonumber
\end{align}
\end{document}